\documentclass[11pt,a4paper]{book}
\usepackage{amsmath}
\usepackage{graphicx}
\usepackage{caption} 
\usepackage{paralist} 
\usepackage{fancyheadings}
\usepackage[%
  colorlinks=true,%
  urlcolor=blue,%
  citecolor=blue%
]{hyperref}
\usepackage[font={small,sl}]{caption} 
\usepackage[authoryear]{natbib} 

\def\cite{\citep}

\newcommand{\figref}[1]{\hyperref[fig-#1]{Figure \ref*{fig-#1}}}
\newcommand{\secref}[1]{\hyperref[sec-#1]{Section \ref*{sec-#1}}}
\newcommand{\todo}[1]{~\textbf{[[ \emph{ TODO #1 } ]]}~}
\newcommand{\remark}[1]{} 

\makeatletter \@beginparpenalty=5000 \makeatother

\newcommand{\AF}{K.\@~Anton~Feenstra}
\newcommand{\SA}{Sanne~Abeln}
\newcommand{\JH}{Jaap~Heringa}

\begin{document}

\frontmatter

\pagestyle{fancy}
\lhead[\small\thepage]{\small\sf\nouppercase\rightmark}
\rhead[\small\sf\nouppercase\leftmark]{\small\thepage}
\newcommand{\innerfoot}{\footnotesize{\sf{\copyright} Abeln \& Feenstra}, 2014-2017}
\newcommand{\outerfoot}{\footnotesize \sf Structural Bioinformatics}
\lfoot[\outerfoot]{\innerfoot}
\cfoot{}
\rfoot[\innerfoot]{\outerfoot}
\renewcommand{\footrulewidth}{\headrulewidth}

\title{Structural Bioinformatics}

\author{\SA \and \AF
  \and
  \\[10ex]
  \textrm{\footnotesize Centre for Integrative Bioinformatics (IBIVU), and }\\
  \textrm{\footnotesize Department of Computer Science, }\\
  \textrm{\footnotesize Vrije Universiteit, De Boelelaan 1081A, 1081 HV Amsterdam, Netherlands}
}

\maketitle

\section*{Abstract}
This chapter deals with approaches for protein three-dimensional structure prediction, starting out from a single input sequence with unknown structure, the `query' or `target' sequence. Both template based and template free modelling techniques are treated, and how resulting structural models may  be selected and refined. We give a concrete flowchart for how to decide which modelling strategy is best suited in particular circumstances, and which steps need to be taken in each strategy. Notably, the ability to locate a suitable structural template by homology or fold recognition is crucial; without this models will be of low quality at best. With a template available, the quality of the query-template alignment crucially determines the model quality. We also discuss how other, courser, experimental data may be incorporated in the modelling process to alleviate the problem of missing template structures. Finally, we discuss measures to predict the quality of models generated.
\newpage

\tableofcontents

\mainmatter

\newcommand{\and}{\quad}

\newpage
\setcounter{chapter}{6}

\chapterauthor{\SA \and \JH \and \AF\\[5ex]
\textrm{\footnotesize Centre for Integrative Bioinformatics (IBIVU) and \\
Department of Computer Science, \\
Vrije Universiteit, De Boelelaan 1081A, 1081 HV Amsterdam, Netherlands}
}
\chapter{Strategies for protein structure model generation}

\begin{figure}
\centerline{\includegraphics[width=0.7\linewidth]{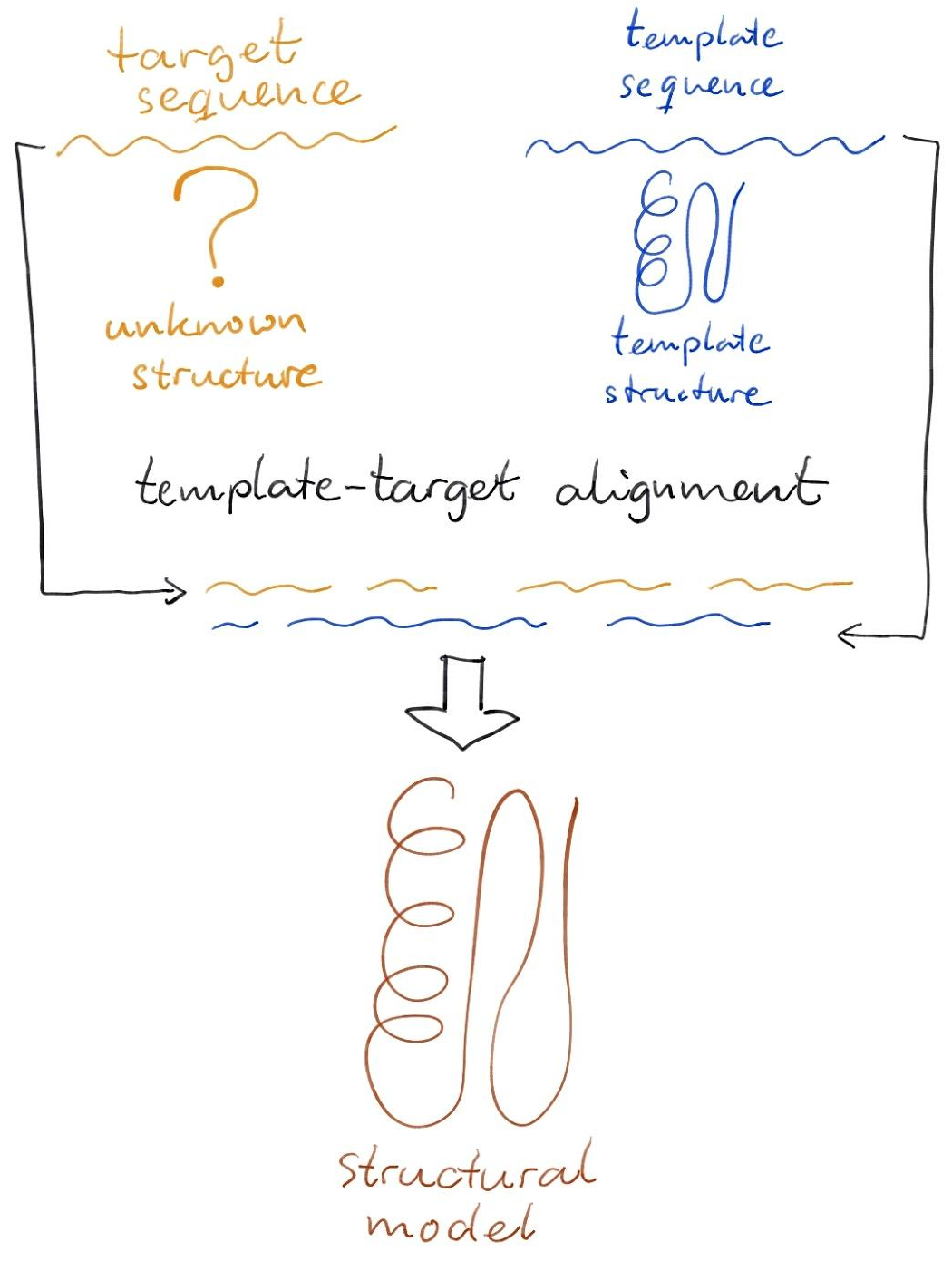}}
\caption{Terminology used in protein structure prediction. We start from our protein of interest (with no known structure): the target sequence. First step is find a matching protein: a template sequence with known structure; the template structure. We then create a template-target sequence alignment, and from this alignment create the structural model which is the solution structure for our target protein.}
\label{fig-3DPred-Terminology} 
\remark{Copyright OK: from scratch by Anton}
\end{figure}

\begin{figure}
\vspace*{-50pt}%
\includegraphics[width=\linewidth]{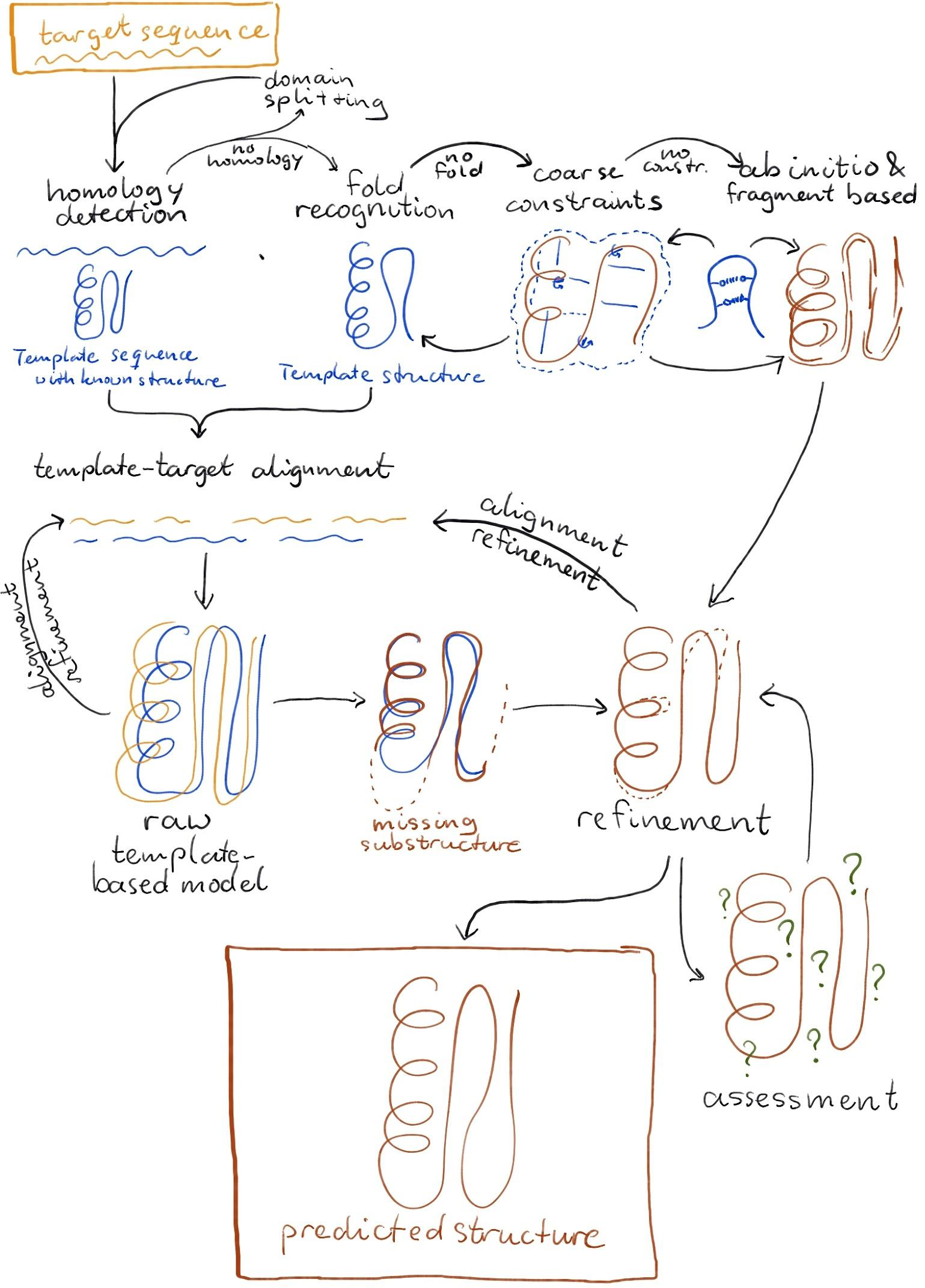}
\captionsetup{width=1.25\linewidth}
\caption{Flowchart of protein three-dimensional structure prediction. It starts at the top left with the target protein sequence of interest, and ends with a predicted 3D structure at the bottom. Depending on the availability of a homologous template, a suitable fold, or coarse/experimental constraints, different options are available, with sharply decreasing expected model accuracy for each step. Homology-based models (far left) are most accurate, while ab-initio modelling (far right) is notoriously unreliable. Template-based (by homology or fold recognition) models require an alignment with the target sequence, from which the initial model will be built. Course constraints may sometimes be incorporated in this stage. The raw model may need to be completed by separate modelling of the missing substructures. Template-free modelling can benefit greatly if such constraints are available, if not the only sources left are fragment libraries and knowledge-based energy functions. The model, whether template-based or -free, is usually refined until a desired level of (estimated) quality is reached to produce the final predicted structure for our target protein sequence of interest.} 
\label{fig-3DPred-Flowchart}
\vspace*{-20pt}%
\end{figure}

Here we consider strategies for a typical protein structure prediction problem: we want to generate a structural model for a protein with a sequence, but without an experimentally determined structure.  
In the previous chapter on ``Introduction to protein structure prediction'' we introduced the problem of how to obtain the folded structure of the protein, given only an amino acid sequence.
In this chapter, we will build up a workflow for tackling this problem, starting from the easy options that, if applicable, are likely to generate a good structural model, and gradually working up to the more hypothetical options whose results are much more uncertain. 
We will in detail discuss first template-based and then template-free modelling. An overview of protein structure modelling, including both template-based and template-free modelling is given in \figref{3DPred-Flowchart}; see also \figref{3DPred-Terminology} for the the terminology used.

\section{Template based protein structure modelling}

\subsection{Homology based Template Finding}

Homology modelling is a type of template based modelling with a template that is homologous to the target protein. As mentioned before, homology modelling works well, because structure is more conserved than sequence. Typically, we will use a sequence-based homology detection method, such as BLAST, to search for homologous protein sequences in the full PDB dataset. If we find a sequence that has significant sequence similarity to the full length of our target sequence we have found a template. 
Of course it is possible that a template only covers part of the target sequence, see also the section on domains in Chapter 6 ``Introduction to protein structure prediction''.

If a simple BLAST search against the PDB gives no good results we may need to start using alignment methods that can detect more distant homologues based on sequence comparison, such a PSI-BLAST or HMMs. PSI-BLAST uses hits from a previous iteration of BLAST to create a profile for our query sequence \cite{Altschul1997}. This allows successive iterations to give more weight to very conserved residues, allowing to find more distant homologues. Even more sensitive homology detection tools typically consider sequence profiles of both the query and the template sequence, and try to align / score these profiles against each other. Examples: Profile-Profile Alignment \todo{ref?}, Compass \cite{Sadreyev2003}, HMM's such as HHpred \cite{Soding2005}. Note that, to make full use of such methods it is important that the searching profiles are generated using extended sequence databases (so not just on the PDB). The evolutionary sequence profiles used in these methods can ensure that the most conserved, and therefore structurally most important,  residues are more likely to be aligned accurately. 

Once we have a good template, and an alignment of the target sequence with the template structure, we can build a structural model; this is called homology modelling in cases of clear homology between the target and the template. 


\subsection{Fold recognition}

If no obvious homologs with a known structure can be found in the PDB, it becomes substantially more difficult to predict the structure for a sequence. We may then use another trick to find a template: remember that the template does not only have a sequence, but also a structure. Therefore, we may be able to use structural information in this search. Fold recognition methods look explicitly for a plausible match between our query sequence and a \emph{structure} from a database. Typically, such methods use structural information of the putative template to determine a match between the target sequence and the putative template sequence and structure.

Note that, in contrast to homology-based template search, it is not strictly necessary for a target sequence and a template sequence to be homologous -- they may have obtained similar structures through convergent evolution. Therefore some fold recognition methods are trained and optimised for detection of structural similarity, rather than homology. In fact, for very remote homologs we may not be able to differentiate between convergent and divergent evolution. For the purpose of structure prediction this may not be an important distinction, nevertheless divergent evolution, i.e., homology through a shared common ancestor, invokes the principle of structure being more conserved, giving more confidence in the final model.

Threading is an example of a fold recognition method \cite{Jones1992}. The query sequence is threaded through the proposed template structure. This means structural information of the template can be taken into account when score if the (threaded) alignment between the target and the potential template is a good fit. Typically threading works by scoring the pairs in the structure that make a contact, given the sequence composition. A sequence should also be aligned (threaded) onto the structure giving the best score. Scores are typically based on knowledge based potentials. For example, if two hydrophobic residues are in contact this would give a better score than a contact between a hydrophobic and polar amino acid. Note that threading does not necessarily search for homology. Threading remains a popular fold recognition methods, with several implementations available \cite{Jones1992,Zhang2008,Song2013}.

Another conceptually different fold recognition approach is to consider that amino-acid conservation rates may strongly differ between different structural environments. For example, one would expect residues on the surface to be less conserved, compared to those buried in the core. Similarly, residues in a $\alpha$-helix or $\beta$-strand or typically more conserved than those in loop regions. In  fact, the chance to form an insertion or deletion in a loop regions is seven times more likely than within (other) secondary structure elements. Since we know the structural environment of the residues for the potential template, we can use this to score an alignment between the target sequence and potential template sequence. FUGUE is a method that scores alignments using structural environment-specific substitution matrices and structure-dependent gap penalties \cite{Shi2001}.

\subsection{Generating the target-template alignment}
Once a suitable template has been found, one can start building a structural model. Typical model building methods will need the following inputs: (1) the target sequence, (2) the template structure, (3) sequence alignment between target and template
and (4) any additional known constraints. The output will be a structural model, based on the constraints defined by the template structure and the sequence alignment.

Here, it is important to note that the methods that recognize good potential templates, are not necessarily the methods that will produce the most accurate alignments between the target and template sequence. As the final model will heavily depend upon the alignment used, it is important to consider different methods, including for example structure and profile based methods,  multiple sequence alignment programs or methods that can include structural information of the template in constructing the alignment. Examples of good sequence-based alignment methods, which can also exploit evolutionary signals and profiles, are \textsl{Praline} \cite{Simossis2005} and \textsl{T-coffee} \cite{Notredame2000}. The T-coffee suite also includes the structure-aware alignment method \textsl{3d-coffee} \cite{OSullivan2004}. Some aligners are context-aware, taking into account for example secondary structure \cite{Simossis2005a} or trans-membrane (TM) regions explicitly \cite{Pirovano2008,Pirovano2010a,Floden2016}, which are useful extension as TM proteins are severely underrepresented in the PDB. Of particular interest to the general user are automated template detection tools such as HHpred \cite{Soding2005}.
\citet{Bawono2017} give a good and up-to-date overview of multiple sequence alignment methodology, including profile-based and hidden-Markov-based methods. 

Moreover, once a first model is created, it may be wise to interactively adapt the alignment, based on the resulting model, which might lead to an updated model. This procedure may also be carried out in an interative fashion. 

\subsection{Generating a model}

Here we consider the \textsl{MODELLER} software to generate template based alignments \cite{Sali1993}, which is one of several alternative approaches to construct homology models \cite{Schwede2003,Zhang2008,Song2013}.
Firstly, the known template 3D structures should be aligned with the target sequence. Secondly, spatial features, such as C$\alpha$-C$\alpha$ distances, hydrogen bonds, and mainchain and sidechain dihedral angles, are transferred from the template(s) to the target. \textsl{MODELLER} uses ``knowledge based'' constraints. The constraints are based on the template distances, the alignment, but also on knowledge- based energy functions (probability distribution). The constraints are optimised using molecular dynamics with simulated annealing.  Finally, a 3D model can be generated by satisfying all the restraints as well as possible.

\subsection{Loop or missing substructure modelling}
We now have a model for all the residues that were aligned well between the template and the target. The remaining substructure(s), that are not covered by the template, will show as gaps in the alignment between target sequence and template. `Loop modelling' is used to determine the structure for these missing parts. Loop models are typically based on fragment libraries, knowledge based potentials and constraints from the aligned structure. This problem is in fact closely related to the template-free modelling procedure, as we need to generate a structure without a readily available template (template-free approaches are discussed in more detail in the next section). In CASP11, consistent refinement overall as well as for loop regions was achieved; the limiting factor for effective refinement was concluded to be the energie functions used, in particular missing physicochemical effects and balance of energy terms \cite{Lee2016}.

\section{Template-free protein structure modelling}
\subsection{What if no suitable template exists?}

If on the other hand, no suitable template is available for our target protein of interest, we will need to follow a `template-free' modelling strategy. Without a direct suitable template, we need an ``ab initio'' strategy that can suggest possible structural models based on the sequence of the template alone. In this case, we need to resort to fragment based approaches. Here small, suitable fragments, from various PDB structure, are assembled to generate possible structural models. As the fragments are typically matched using sequence similarity, one may even consider this as template based modelling at a smaller scale. However, since the sequence match is based on a limited number of residues, this would not generally imply a homologous relation between the fragment template and the target sequence. It is also important to note that that this type of ``ab initio'' modelling is still ``knowledge based'': the structural models are generated from small substructures present in the PDB, assessed by energy scoring functions generated by mining the PDB. In other words, these model are not based on physical principles and physico-chemical properties alone. This also means, that such models are likely to share any of the biases that are present in the PDB, such as lack of trans-membrane proteins and absence of disordered regions.

\begin{figure}
\centerline{\includegraphics[width=1.5\linewidth]{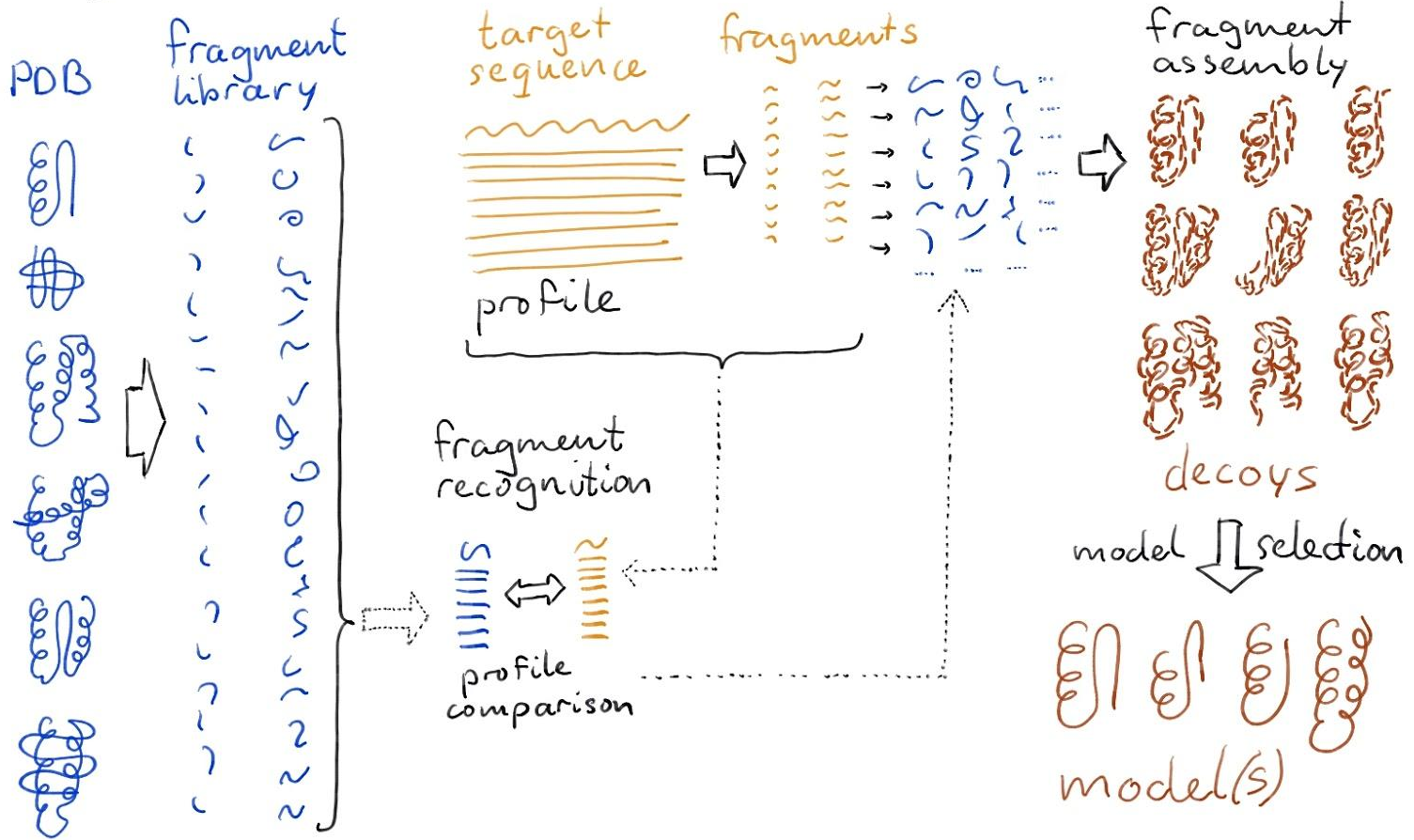}}
\caption{Overview of the fragment-based modelling strategy. A library of structure fragments was created once from the PDB; all small 3-residues and larger 9-residue fragments are collected and clustered. A target sequence of interest is also separated into 3- and 9-residue sequence fragments. For each of these, a profile-profile search is performed to find matching fragments from the fragment library; typically for each target fragment, multiple hits with different structure are retrieved. This collection of fragments of alternate structure are then assembled through a Monte Carlo algorithm into a large set of possible structures, called `decoys'. Using knowledge-based potentials and overall statistics, from the decoy set, a final selection of model structures is made. }
\label{fig-3DPred-fragment} 
\remark{Copyright OK: from scratch by Anton}
\end{figure}

\subsection{Generating models from structural fragments}

Here we follow the keys ideas in the \textsl{Rosetta} based ab initio modelling suite \cite{Simons1999}. Quark, another fragment-based approach provides a similar performing alternative\cite{Xu2012}. The overall approach is to split the target sequence into 3 and 9-residue overlapping sequence fragments, i.e., sliding windows, and find matching structural fragments from PDB. 



A fragment library is generated by taking 3 and 9-residue fragments from the PDB and clustering these together into groups of similar structure. For each fragment in the database sequence profiles are created. These profiles subsequently are used to search for suitable fragments for our query/target sequence, as we only have sequence information (see \figref{3DPred-fragment} on the left-hand side).

The target sequence also will be split into 3 and 9-residue overlapping sequence fragments. These target sequence fragments are then matched with the structural PDB based fragment library using profile-profile matching. Note that this procedure will generate multiple fragments for each fragment window in the sequence. The fragment windows on the sequence typically also overlap (see \figref{3DPred-fragment} middle panel).

\subsection{Fragment Assembly into decoys}


The above scenario leads to many possible structural fragments to cover a sequence position. To find the best structural model is a combinatorial problem in terms of fragment combinations, see \figref{3DPred-fragment} top right. A Monte Carlo algorithm is used to search through different fragment combinations. Good combinations are those that give a low energy. Each MC run will produce a different model, since it is a stochastic algorithm. Note that to be able to optimise a model, we need a scoring function. A knowledge based energy function is used, including the number of neighbours, given amino acid type, residue pair interactions, backbone hydrogen bonding, strand arrangement, helix packing, radius of gyration, Van der Waals repulsion. These are all terms that are relatively cheap to compute when a new combination of fragments is tried. The structural model is optimised by slowly lowering the MC temperature -- this is also called simulated annealing. The generated models are called `decoys'.

Thereafter, decoys are refined using additional Monte Carlo cycles, and a more fine-grained energy function including: backbone torsion angles, Lennard-Jones interactions, main chain and side-chain hydrogen bonding, solvation energy, rotamers and a  comparison to unfolded state. 

Finally, the most difficult task is to select, from all the refined decoys, a structure that is a suitable model for the target sequence, see \figref{3DPred-fragment} bottom right. Again, using a more detailed, knowledge-based energy function, decoys can be scored to assess how `protein-like' they are. Such a selection procedure may get rid of very wrong models. However, selecting the best model, without any additional information (from for example experiments or co-evolution-based contact-prediction), is likely to lead to poor results (see Chapter 6 'Introduction to Structure Prediction').

\subsection{Constraints from co-evolution based contact prediction or experiments}

As already mentioned, valuable additions to the modelling process are coarse constraints from experimental data or contact prediction. Experimental data from NMR and chemical cross-linking can yield distance restraints that are particularly useful in the template-free modelling to narrow down the conformational space to be searched; still average accuracy of models produced remains extremely limited \cite{Kinch2016b}. Other sources of information are contours or surfaces that can be obtained from cryo-EM, or small angle scattering experiments, either with electrons, neutrons, or x-ray radiation. However, since these techniques are employed mostly for elucidating larger macromolecular complexes, they are considered out of scope for the current chapter.

Of more general applicability may be methods for predicting intra-protein residue contacts; the main approach currently is based on some form of co-evolution information obtained by direct-coupling methods from 'deep' alignments \cite{Marks2011a,Jones2012,Morcos2011}. Depth here signifies the amount of sequence variation present in the alignment in relation to the length of the protein (the longer the protein, the more variation is needed). \citet{Ovchinnikov2017a} expresses this as the \emph{effective protein length}: $Nf = {N80\%ID}/{\sqrt{l}}$ where $l$ is the protein length, and $N80\%ID$ the number of cluster at 80\% sequence identity. They showed that $Nf$ can be greatly enhanced by the use of metagenomic sequencing data, and that this leads to a marked improvement in model quality, and estimate that this would triple the number of protein families for which the correct fold might be predicted \cite{Ovchinnikov2017a}. \citet{Wuyun2016} investigated `consensus'-based methods, which combine both direct-coupling and machine-learning approaches, and find that the machine-learning methods are less sensitive to alignment depth and target difficulty, which are crucial factors for success for the direct-coupling methods.

\section{Selecting and refining models from structure prediction}
Once we have created (several) models, we need to assess which model is the best one. Typically this can be done by scoring models on several properties using model quality assessment programs and visual inspection with respect to ``protein like'' features. Moreover, if any additional knowledge about the structure or function of our target protein is available, this may also help to assess the quality of the model(s). In addition, one may in some cases want to improve a model, or parts of it; this is called model refinement.

\subsection{Model refinement}

For many years in CASP, model refinement was a no-go area; the rule of thumb was: build our homology model and do not touch it! An impressive example of the failure of refinement methods was shown by the David Shaw group, who concluded that ``simulations initiated from homology models drift away from the native structure'' \cite{Raval2012}. 
Since CASP10 in 2014 \cite{Nugent2014} and continuing in the latest CASP11, there is reason for moderate optimism. General refinement strategies report small but significant improvements of 3-5\% over 70\% of models \cite{Modi2016b}. Interestingly, and in stark contrast to the earlier results by \cite{Raval2012}, the average improvement of GDT\_HA using simulation-based refinement now also is about 3.8, with an improvement (more than 0.5) for 26 models \cite{Feig2016}. For five models, the scores became worse (by more than 0.5), and another five showed no significant change. Particularly, for very good initial models (GDT\_HA$> 65$), models were made worse. Moreover, they also convincingly showed that both more and longer simulations consistently improved these results%
; note however that protocol details such as using C$\alpha$ restraints, are thought to be the limiting factor \cite{Feig2016}, as already used previously \cite[e.g.,][]{Keizers2005,Feenstra2006}, and replicated by others \cite[e.g.,][]{Cheng2017}. Most successful refinement appears to come from correctly placing $\beta$-sheet or coil regions at the termini \cite{Modi2016b}.

\subsection{Model quality assessment strategies}


It may be generally helpful to compare models generated by different prediction methods; if models from different methods look alike (more precisely if the pair has a low RMSD and/or high GDT\_TS) they are more likely to be correct. Similarly, templates found by different template finding strategies, found for example both by homology sequence searches and fold recognition methods are more likely to yield good modelling results \cite{Moult2016b,Kryshtafovych2016}. Such a consensus template is generally more reliable than the predictions from individual methods -- especially if the individual scores are barely significant. Lastly, one can consider biological context to select good models.

Whether a model is built using homology modelling, fold recognition and modelling or \emph{ab initio} prediction, all models can be given to a Model Quality Assessment Program (MQAP) for  model validation.  A validation program provides a score predicting how reliable the model is. These scores typically take into account to what extent a model resembles a ``true'' protein structure. The best performing validation programs take a large set of predicted models, and indicate which out of these is expected to be the most reliable. 

Validation scoring may be based on similar ideas as validation for experimental structures or may be specific to structure prediction. For example, it can be checked if the amount of secondary structure, e.g. helix and strand vs. loop, has a similar ratio as in known protein structures; if a model for a sequence of 200 amino acids does not contain a single helix or $\beta$-strand, the model does not resemble true protein structures, and is therefore very unlikely to be the true structural solution for the sequence. 
A similar type of check may be done for the amount of buried hydrophobic groups and globularity of the protein. 

Different models may also be compared to each other. One trick that is commonly used, is that if multiple prediction methods create structurally similar models, these models are more likely to be correct. Hence, a good prediction strategy is to use several prediction methods, and pick out the most consistent solution. A pitfall here is that if all models are based on the same, or very similar, templates, they will look similar but this may not indicate the likelihood of them being correct.

\subsection{Secondary Structure Prediction}

Secondary structure prediction is relatively accurate \cite[see e.g.\@ the review by ][]{Pirovano2010}. This problem is in fact much easier to solve than three-dimensional structure prediction, as is shown in Chapter{ \todo{ref to other chapter in book -- presumably}}. The accuracy of assigning strand, helix or loops to a certain residue can go up to 80\% with the most reliable methods. Typically such methods use (hydrophobic) periodicity in the sequence combined with phi and psi angle preferences of certain amino acid types to come to accurate predictions. The real challenge lies in assembling the secondary structure element in a correct topology. Nevertheless, secondary structure prediction may be used to assess the quality of a model built with a (tertiary) structure prediction method. Many (automated) methods also incorporate secondary structure information during alignment \cite{Simossis2005a}, homology detection \cite{Soding2005,Shi2001} and contact prediction \cite{Terashi2017,Wang2017}.

\bibliographystyle{apalike}
{\small\raggedright
\bibliography{prot3dpred}
}

\end{document}